# The Discussion on Shannon channel capacity formula from the viewpoint of signal uncertainty and Research on the Technique of Breaking through the Shannon Limit

Dequn Liang* and Xinyu Dou

Information Science and Technology College, Dalian Maritime University, Dalian, China

Corresponding author: Dequn Liang. Email: ldqlgn@163.com

**Abstract**

In this paper, firstly, the Shannon channel capacity formula is briefly stated, and the relationship between the formula and the **signal uncertainty principle** is analyzed in order to prepare for deriving the formula which is able to **break through the Shannon channel capacity**. Then, as a practical example of breaking the Shannon limit, the time-shift non orthogonal multicarrier modulation technology is introduced. After more than twenty years of development, this technique is proved to be a practical modulation technique for digital communication.

## 1. Introduction

In 1948, the article "A mathematical theory of communication" [1] by Shannon answered the question of the source coding, channel coding and the channel capacity, laying the foundation of signal processing and communication techniques. Meanwhile, Shannon's information theory plays an important role in many other disciplines such as economics and sociology. As pointed by Thomas M. Cover in his book *Elements of Information Theory*, "...For this reason some consider information theory to be a subset of communication theory. We argue that it is much more. Indeed, it has fundamental contributions to make in statistical physics (thermodynamics), computer science (Kolmogorov complexity or algorithmic complexity), statistical inference (Occam's Razor: "The simplest explanation is best"), and to probability and statistics (error exponents for optimal hypothesis testing and estimation)."[2]

The **channel capacity formula** in Shannon's information theory defined the upper limit of the information transmission rate under the additive noise channel. So far, the communication technique has been rapidly developed to approach this theoretical limit.

Nowadays, the fifth-generation mobile communication technology (5G) has been commercially deployed. Through the mobile communication technique, smart devices have been integrated to the Internet more and more deeply, which implies that a new industrial revolution is coming soon. Due to the unprecedented rising demand of flexible, efficient and reliable communication between people, people and objects even between objects, the wireless communication technique must provide higher data transmission rate, higher spectral efficiency and lower delay than ever [3-5]. For this reason, many technical solutions have been proposed in the industry, such as expanding the spectral resources, developing various new multi-user accessing technologies, signal processing technologies and coding methods, etc. [6-18] At the same time, massive MIMO will be deployed in 5G, and the data transmission rate can be further enhanced by dramatically increasing the number of antennas [7, 8, 19-23]. However, at present, the study of the communication technology focuses merely on the coding techniques to approximate the Shannon Limit, and the research on the fundamental modulation technique which can approach or even break through the Shannon Limit is rarely reported. Moreover, although it can increase the communication capacity, the nature of MIMO is multiplexing multiple spatial physical channels and the Shannon Limit of one single physical channel is still not broken through.

In this paper, firstly, the Shannon channel capacity formula is briefly stated, and the theoretical completeness of the formula is analyzed based on the **signal uncertainty principle**. Then, according to the information entropy theory introduced by Shannon, a theoretical method is developed to break through the Shannon Limit. This method is called **the time-delay overlapping method.** At the end of the paper, we combine this method with a kind of digital modulation technology that has been researched continuously in the past 20 years, which is named as Time Shift- Non orthogonal Multi carrier Modulation Technology (TS-NMT), and verify that the TS-NMT is a practical digital modulation technology that can break through the Shannon limit.

## 2. Channel Capacity of the continuous channel under perturbation

In communication, the information source can be divided into the discrete source and the continuous source. The former is to transmit symbols discontinuously, e.g. Morse, and the latter is to transmit continuous physical variables such as electricity, light or underwater acoustics, etc. Similarly, the channel can be divided into the discrete and continuous channel. Generally, in the communication industry, the discussion of the Shannon Limit aims at the capacity of the continuous channel.

A large amount of mathematical descriptions were employed by Shannon to build the information theory. However, in this paper, it is impossible to repeat those mathematical descriptions, so we directly quote some expressions under the acknowledgment of the correctness of Shannon's mathematical descriptions and insert necessary explanations for easily comprehending.

Science requires exact and quantitative expressions. Nevertheless, to interpret the uncertainty principle in quantum mechanics from the perspective of philosophy, we can say that it is impossible for human to achieve the ideal exact and quantitative expressions. Therefore, in reality people must find adequate methods from different aspects to deal with that contradiction. In communication, the uncertainty is an important adverse factor caused by various perturbations (without losing of generally, we only use the noise to represent the perturbation in the rest of this paper). Thoughts and methods of probability theory can be used to handle the uncertainty problem. Specifically, through making the average (or expectation) of all the probabilities in the set of possible events, the uncertainty problem will transform into a measurable problem, and finally, we can solve the uncertainty problem through certain methods.

Firstly, we should distinguish two words named message and information. The message represents the occurred event. The common sense tells us that a message representing small probability events will cause more attention of the receiver than a high probability event. To quantitatively interpret this phenomenon, let $x_i$ represent a message, and the amount of information of the message is $I(x_i) = \log \frac{1}{p(x_i)} = -\log p(x_i)$, where $p(x_i)$ is the probability density. From this expression, we can say that the amount of information of the message representing the small probability event is large -- and vice versa. Hartley was the first person to introduce this definition. [24] In the communication domain, there is $I(x_i) = \log_2 \frac{1}{p(x_i)} = -\log_2 p(x_i)$, where base 2 makes the unit of $I(x_i)$ binary digits or bits (In Shannon's papers base 2 was always omitted). The randomness of message is one of the important reasons for the uncertainty of communication system, and information quantity $I(x_i)$ can quantify the uncertainty of message (random variable) $x_i$, which can be used to quantitatively describe the "uncertainty" problem in communication. The signals $x_i$ and $y_i$ at both ends of the channel form a unified probability space. The change of two random events will affect each other's information quantity. Therefore, the mutual information quantity should also be defined:

In the receiver, we have $I(x_i;y_i)=\left[-\log p(x_i)\right]-\left[-\log p(x_i/y_i)\right]=\log\dfrac{p(x_i/y_i)}{p(x_i)}$, where $-\log p(x_i)=\log\dfrac{1}{p(x_i)}$ is the exact amount of information of the transmitted signal $x_i$, and $-\log p(x_i/y_i)=\log\dfrac{1}{p(x_i/y_i)}$ implies that in $y_i$ a partial of the information about $x_i$ is ambiguous and becomes the uncertain information due to the noise in the channel. In order to obtain the exact received information, the uncertain part of the information must be subtracted from the whole received information.

In the transmitter, we have $I(y_i;x_i)=\left[-\log p(y_i)\right]-\left[-\log p(y_i/x_i)\right]=\log\dfrac{p(y_i/x_i)}{p(y_i)}$, which implies that to forecast the exact information in the receiver, the uncertain information $-\log p(y_i/x_i)$ caused by the noise should be subtracted from $-\log p(y_i)$.

Finally, we have the mutual information

$$I(X;Y)=\sum_{i=1}^{n}\sum_{j=1}^{m}p(x_iy_j)I(x_i;y_j)=\sum_{i=1}^{n}\sum_{j=1}^{m}p(x_iy_j)\log\dfrac{p(x_i/y_j)}{p(x_i)}$$
$$=\sum_{i=1}^{n}\sum_{j=1}^{m}p(x_iy_j)\log\dfrac{p(y_j/x_i)}{p(y_j)}=\sum_{i=1}^{n}\sum_{j=1}^{m}p(x_iy_j)I(y_j;x_i)=I(Y;X)$$
(1)

Shannon considered the whole possible values of the random variable, i.e. the expectation of the amount of information of all the possible events, and introduced the definition **Entropy** in the mechanics of machinery discipline to further quantify the uncertainty of the random variable. The amount of information of the discrete source containing $M$ symbols is defined as the statistical average of the amount of information of each symbol, and can be expressed as

$$H(X)=-\sum_{i=1}^{N}p(x_i)\log p(x_i) \qquad (2)$$

where $X$ and $H(X)$ represents the random variable and the entropy, respectively. Obviously, Eq. (2) is the statistical average of the amount of information of a random variable. Provided that $p(x_i)$ is known for all $x_i$ (the probability theory gives many methods to achieve this), $H(X)$ has an exact value and the uncertainty of $X$ is therefore quantified.

For the continuous source, the value of the continuous physical variable is divided into $2M$ sections, i.e. $\Delta x_i, i=-N,...,-1,1,...,N$, and the probability of the value of the physical variable falls in $\Delta x_i$ is $p(x_i)\Delta x_i$. Then the probability density of the continuous random variable is $p(x) = \lim_{\Delta x_i \to 0} p(x_i) \Delta x_i$. Thus, the amount of information of the continuous source is defined as

$$H(X) = -\int_{-\infty}^{\infty} p(x) \log p(x) dx \tag{3}$$

When $p(x) = \frac{1}{\sigma\sqrt{2\pi}} e^{-x^2/2\sigma^2}$ and $\int_{-\infty}^{\infty} p(x) dx = 1$, $H(X)$ achieves the maximum, and the uncertainty of random variable can be finally quantified by used $H(X)$. At this time the $p(x)$ is defined as the best probability density distribution.

Let the source of the communication system with perturbed channel be a continuous source, and let $x(t)$ be the transmitted signal, whose power is $P$. At this time, the received signal is also a continuous signal and expressed as $y(t) = x(t) + n(t)$, where $n(t)$ represents the noise. The time variable $t$ indicates the continuous of the signal in time domain. According to the sampling theory, for a signal with bandwidth $W$ and period $T$, we can use no less than $2TW$ sampling points to fully represent it. We will firstly discuss the amount of information of a single sampling point (in this scenario, the time variable $t$ can be omitted), and signals at two sides of the continuous channel with perturbation can therefore be simply represented as $x$ and $y=x+n$. Now we have $H(X)$, $H(N)$ and $H(Y)=H(X+N)$ representing the entropy of the transmitted signal, the noise and the received signal, respectively. According to the property of additivity of the entropy, only if the transmitted signal $x$ and the noise $n$ are statistically independent we have $H(Y) = H(X) + H(N)$ or $H(X) = H(Y) - H(N)$. This indicates that the average amount of information of the received signal contains that of the noise. That is to say, the noise makes a portion of the signal ambiguous and enhance the uncertainty of the system. This uncertainty can be expressed as $H(N)$ or the conditional entropy $H(X/Y)$ (also called the equivocation). In another way, if there is no noise, we will have $H(Y)=H(X)$, which is absolutely certain. The uncertainty is caused by the noise, which can be represented by the conditional entropy $H(X/Y)$, and therefore we can get the exact amount of information of the received signal through subtracting the $H(X/Y)$, i.e. $H(X)-H(X/Y) = H(Y) - H(Y/X) = H(Y) - H(N)$. At this time, the mutual information can be expressed as $I(X;Y) = H(X) - H(X/Y)$. According to Eq. (1), we can get the relationship between the mutual information and the aforementioned entropies as follows

$$I(X;Y) = H(X) - H(X/Y) = H(Y) - H(Y/X) = H(Y) - H(N) \tag{4}$$

The least $2TW$ sampling points given by the sampling theory can ensure the independent of each sampling point and the same probability density function. Thus, the statistical average of the amount of information of $2TW$ discrete sampling points in

a cycle is the transmission rate of the channel with perturbation and can be expressed as

$$R = I(X;Y) \cdot 2TW/T = [H(Y) - H(N)] \cdot 2TW/T \tag{5}$$

Shannon defined the maximum transmission rate as the channel capacity

$$C = R_{max} = \max\{I(X;Y) \cdot 2W\} \tag{6}$$

Shannon had already proven that under the condition of the best probability density distribution and additive Gaussian white noise, the entropies of the transmitted and received signal achieve the maximum, i.e. $H(Y) = \log_2 \sqrt{2\pi(P+N)}$ and $H(N) = \log_2 \sqrt{2\pi\sigma^2} = \log_2 \sqrt{2\pi N}$, where $N$ is the power of the noise. By putting all the above equation into Eq. (5) and Eq. (6) we can get the Shannon Limit formula as follows

$$C = W \log_2 \frac{P+N}{N} = W \log_2 \left(1 + \frac{P}{N}\right) = W \log_2 (1 + SNR) \tag{7}$$

Or it can be expressed as the spectral efficiency form as

$$C/W = \log_2 \frac{P+N}{N} \tag{7-1}$$

where $P$ and $N$ represent the power of the signal and the noise, respectively, and *SNR* represents the signal-to-noise ratio. Eq. (7) and Eq. (7-1) are defined as the Shannon channel capacity formula and are called as the Shannon Limit in the industry. That is to say, for a specific *SNR*, the Eq. (7) gives the upper limit of the information transmission rate.

Shannon then proposed a method to approximate this rate limit. Firstly, the statistical property of the transmitted signal must approximate a white noise. Secondly, the least R.M.S error theory can be used to implement the demodulation, i.e. by comparing all the possible $M=2^s$ samples of the transmitted signal with the received signal, the sample which has the least R.M.S. discrepancy from the received signal is chosen as the demodulated signal. This method can be represented as

$$\lim_{\varepsilon \to 0} \lim_{T \to \infty} \frac{\log M(\varepsilon, T)}{T} = W \log \frac{P+N}{N} \tag{8}$$

where $\varepsilon$ and $T$ represent the frequency of errors and the duration of the signal, respectively. This means that no matter how small the $\varepsilon$ is, we can make the transmission rate approximate to $TW\log(1+P/N)$ as possible as we can through sufficiently large $T$. What's more, we must say that, to approximate the transmission rate calculated by Eq. (7), both $\varepsilon \to 0$ and $T \to \infty$ have to be simultaneously satisfied. This just fits the **signal uncertainty principle**. Specifically, only when $T \to \infty$ the bandwidth $W$ of the signal is finite, and the sampling principle employed during deriving Eq. (7) is valid. The signal uncertainty principle is the representation of the uncertainty principle of quantum mechanics in the signal processing domain. As Wang Peng and Li Jian-ping indicated in their article "Quantum Interpretation of Signal's Uncertainty Principle", according to the uncertainty principle of the quantum mechanics, the signal in some sense can be treated as a quantum-like system with wave-particle duality, and just due to the wave-particle duality, we cannot simultaneously determine both the duration and the bandwidth of the signal [25]. In addition, some papers demonstrated the signal uncertainty principle from the Fourier transform point of view [26, 27].

Although Eq. (8) ensures the mathematical completeness of the Shannon Limit formula, there is a problem about the realization in the real word. Firstly, in practice, the infinity in time domain is impossible to be satisfied, and therefore the finite bandwidth in frequency domain is hard to be fulfilled. Fortunately, in the process of project implementation, the bandwidth can be limited under a given frequency of error $\varepsilon$, because various kinds of bandwidth are defined in the industry, such as the 3dB bandwidth, null-to-null bandwidth, minimum fading bandwidth, etc. Then we will face another problem of how to ensure $\varepsilon \to 0$. Gratifying is that Shannon had given the coding theorem as follows. If a source has a rate $R_1$ for a valuation $v_1$ it is possible to encode the output of the source and transmit it over a channel of capacity $C$ with fidelity as near $v_1$ as desired provided $R_1 < C$. This is not possible if $R_1 \geq C$.

Shannon presented the spectral efficiency curve, i.e. the $C/W$ curve, corresponding to Eq. (7-1) in another article [28]. According to the method given by Shannon to approximate the maximum transmission rate, we designed the numerical simulation program and the simulation results are shown in Fig. 1. To save the space of this paper, we do not present the detailed program and just show the main parameters as follows. Let $n_0$ and $n_s$ be the power spectral density (PSD) of the narrow-band Gaussian white noise and the transmitted signal which approximates the white noise, respectively, and $n_s/n_0$ represents the *SNR*. We can get different *SNR* when changing the $n_s$ and meanwhile maintaining the $n_0$. At the same time, both the bandwidth of the transmitted signal and the noise are set to be $W$ (in the program it is set to be 50 kHz) and we have $P_0 = \sigma_0^2 = W n_0$ and $P_{si} = \sigma_{si}^2 = W n_{si}$. In addition, according to the sampling theory, in the simulation the transmitted signal $x(t)$, the noise $n(t)$ and the received signal $y(t) = x(t) + n(t)$ are represented as $x(t_i)$, $n(t_i)$ and $y(t_i) = x(t_i) + n(t_i)$, where i=1,…, $2TW$ and $T$ is the duration of the signal. The demodulation progress is shown in Fig. 1(a). Obviously, in the simulation *T* is impossible to be infinitely large, and we have to repeat the program many times to ensure the simulation results approximate the theoretical formula. The simulation results of the spectral efficiency are presented in Fig. 1(b), and we can see that the simulation results are fitted with the theoretical curve [32].

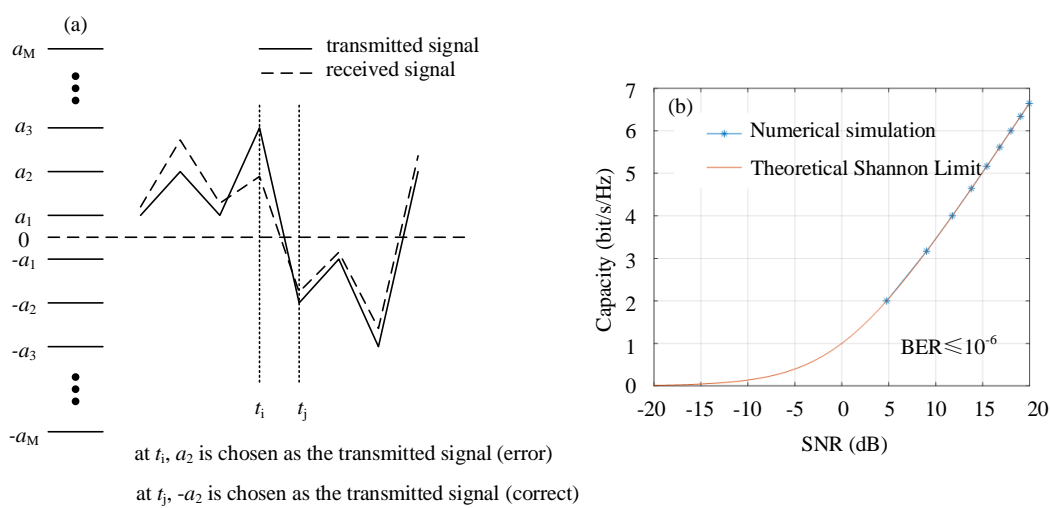

**Figure 1**. Numerical simulation of the Shannon Limit. (a) Demodulation process proposed by Shannon. (b) Theoretical curve of the Shannon Limit and the numerical simulation results.

Looking comprehensively at the formulas (7), (8) and coding theorem given by Shannon, we can say that Shannon established a complete channel theory with mathematics and physics for us 70 years ago! Looking back on Shannon's contribution 70 years later, how can we not admire its greatness! From quantum physicists, we can see Shannon's greatness in another way.

The entropy was firstly employed to describe the change of the temperature under thermal injection by German Physicist R. Clausius in 1865 [29], and the statistical physical expression was proposed by L. Boltzmann in 1877 [30]. Until it was used to measure the information, people really realized that the entropy is a measurement of the disorder of the system. For this reason, E. Jaynes regarded the entropy in the thermodynamics an application of Shannon's information theory [31]. With the development of astrophysics and micro-physics, uncertainty principle proposed by W. Heisenberg in 1927 has become more and more universal [32]. The information theory is therefore no longer limited to the engineering theory but gradually becomes a basic theory, which has a solid mathematical support and can be used in many disciplines.

We should not only praise Shannon's contribution, but also enrich and develop his thoughts and methods. This is also the ultimate goal of this paper.

## 3. Theoretical derivation of Breaking through the Shannon Limit

The technique we proposed here is called the **time-delay overlapping** method. Just by replicating the discrete sampling points of the continuous signal on the base of the derivation of the Shannon Limit in the previous section, and then inserting the newly obtained $2TW$ sampling points into the interval of the origin $2TW$ sampling points in turn, we can get the synthetic signal and then obtain the channel capacity formula with nearly 2 times of the Shannon Limit. The intuitionistic display of the proposed method is shown in Fig. 2.

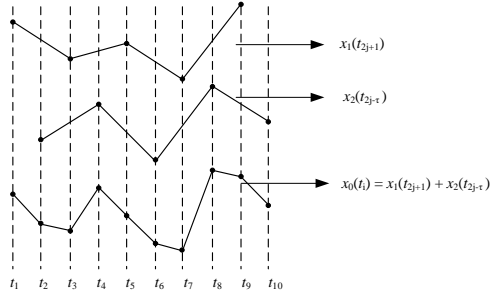

**Fig. 2.** Diagram of the proposed time-delay overlapping method.

We will demonstrate the time-delay overlapping method more clearly using the following simple formulas. Let two continuous signals be $x_1(t_{2j+1})$ and $x_2(t_{2j}-\tau)$, where $j=0, 1, \ldots, 2W-1$ and $\tau=\frac{1}{2}\frac{1}{2W}$. This indicates that the sampling points of the second signal delays half of the sampling interval comparing with the first signal, which means that the sampling points of the second signal are inserted into intervals of sampling points of the first signal in turn. It is specified that the two signals have the same characteristics in terms of bandwidth, PSD and average power, and the sampling points of the two signals are independent of each other[1*]. The corresponding random variables are denoted as $X_1$ and $X_2$, and the entropies are $H(X_1)$ and $H(X_2)$. The synthetic signal can be represented as follows

$$x_0(t_i) = x_1(t_{2j+1}) + x_2(t_{2j}-\tau) \qquad (9)$$

where $i=1, \ldots, 4TW$, $j=0, 1, \ldots, 2TW-1$ and $x_1(t_{4TW}) = x_2(t_0) = 0$. It should be noted that according to the linear characteristics of the Fourier transform, the bandwidth of

---

[1*] When two series of sampling points constitute a single signal and transmit through the channel, neighboring sampling points will experience mutual interference. As long as deploying "ideal" channel estimation and equilibrium, independent sampling points will be obtained in the receiver. Certainly this is the ideal condition and in practice there will be errors. It is the problem of the project implementation.

the synthetic wave composed of $x_1(t_{2j+1})$ and $x_2(t_{2j}-\tau)$ is same with any $x_1(t_{2j+1})$ or $x_2(t_{2j}-\tau)$, that is to say, the bandwidth of $x_0(t_i)$, $x_1(t_{2j+1})$ and $x_2(t_{2j}-\tau)$ is equal. Figure 4(c) in Section 4 is an example of a spectrum composed of sine waves with different time delays and the same frequency. It can be seen that the bandwidth is constant. The entropy of the synthetic signal is $H(X_0)$. According to Eq. (8), the signal uncertainty principle is valid only if i and j are both infinitely large or infinite loop of $T$ is carried out in simulation. However, in the simulation process, only limited loop of $T$ can be implemented and therefore the simulation results are finite and approximate.

In the receiver, the signal is expressed as $y_0(t_i)= x_0(t_i) + n_0(t_i) = x_1(t_{2j+1}) + x_2(t_{2j}-\tau) + n_0(t_i)$, where $n_0(t_i) = \begin{cases} n_1(t_{2i+1}) \\ n_2(t_{2i}) \end{cases}, i=1,...,4TW$. The mean power of the noise is

$$N_0 = \left\{\left[n_1(t_{2i+1})\right]^2 + \left[n_2(t_{2i})\right]^2\right\}/4TW = \left[n_1(t_{2i+1})\right]^2/2TW = N \qquad (10)$$

This indicates that for the proposed time-delay overlapping method, the power of the noise is not changed.

According to the reciprocity theorem, we rewrite the Eq. (4) and Eq. (5) as

$$I(X_0;Y_0) = H(X_0) - H\left(\frac{X_0}{Y_0}\right) = H(Y_0) - H\left(\frac{Y_0}{X_0}\right) = H(Y_0) - H(N) \qquad (11)$$

and

$$R_0 = I(X_0;Y_0) \cdot \frac{4TW}{T} = \left[H(X_0) - H\left(\frac{X_0}{Y_0}\right)\right] \cdot 4W = \left[H(Y_0) - H\left(\frac{Y_0}{X_0}\right)\right] \cdot 4W \qquad (12)$$

Thus, the channel capacity of the time-delay overlapping method is

$$C_0 = R_{0\max=}I(X_0;Y_0) \cdot 4W \qquad (13)$$

Due to the same probability density distribution of $X_0$, $Y_0$ and $X$, $Y$, we can get

$$C_0 = \max\left[H(Y_0) - H\left(\frac{Y_0}{X_0}\right)\right] \cdot 4W = 2W \log \frac{P_0 + N_0}{N_0} \qquad (14)$$

The comparison between Eq. (7) and Eq. (14) should be under the condition of the same power of the signal and the noise. Due to the time delay and nonlinearity of the signal, in the construction process of the synthesized signal, the positive and negative amplitudes are interleaved, and cancellation of the amplitudes will occur in many time periods, so that the power of the synthetic signal will be decreased. We can express the power as $P_0 = \alpha P \leq P_1 + P_2$. The $\alpha$ can be determined by numerical simulation and we can make sure that the scope of $\alpha$ is $0 \leq \alpha \leq 2$. We call this phenomenon the **power**

**efficiency** and will further demonstrate this phenomenon in the next section. This phenomenon means that, by using the time-delay overlapping method, fewer power of the synthetic signal can be consumed when carried data bits of one symbol is unchanged. Moreover, we have proved $N_0=N$. For this reason, we can rewrite Eq. (14) as

$$C_0 = \max\left[H(Y_0) - H\left(\frac{Y_0}{X_0}\right)\right] \cdot 4W = 2W \log \frac{\alpha P + N_0}{N_0} = 2W \log(1 + \alpha P/N) \quad (15)$$

or

$$C_0/W = 2\log(1 + \alpha P/N) \quad (16)$$

$C_0 > C$ indicates that the Shannon Limit has been broken through. The above derivation process can easily be expanded to the situation where there are $L$ independent signals with identical probability density function as follows

$$C_L = LW \log \frac{P_L + N}{N} = LW \log \frac{\beta P + N}{N} = LW \log(1 + \beta P/N) \quad (17)$$

or

$$C_L/W = L\log(1 + \beta P/N) \quad (18)$$

where $P_L = \sum_{l=1}^{L} P_l$ and $0 < \beta \leq L$。

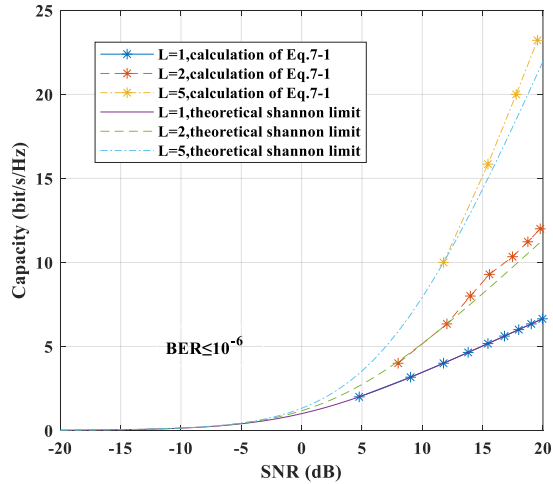

**Fig. 3.** Channel capacities based on the time-delay overlapping method.

The above process describes a method to break through the Shannon Limit. In order to distinguish, we call the channel capacity described by Eq. (7) and Eq. (17) the Shannon Capacity (ShC) and the Time-delay overlapping Shannon Capacity (TDShC). The simulation results of the channel capacity corresponding to Eq. (18) are shown in Fig. 3 under three different conditions, i.e. $L = 1$, 2, and 5. We can draw some conclusions as follows. Firstly, the TDShC is larger than ShC when $L$ is bigger than 1, which means that the Shannon Limit has been broken through. Secondly, when $L$ increases, the TDShC is obviously enhanced, which indicates that large $L$ will bring a higher channel capacity.

## 4. An example to break through the Shannon Limit -- Digital communication technology based on the time-delay overlapping modulation method

In 1997, authors started to explore a more efficient modulation method for digital communication. In 2000, the patent "Phase-shift overlap codes and decode method " (grant no. 1151641) is authorized. Then, authors implemented lots of theoretical studies, numerical simulations and physical experiments, and acquired many new patents. Among those patents, the patent "Time-frequency-phase mixed multicarrier modulation method" (grant no. 101662437) reflect more comprehensively the idea of constructing high-efficient digital modulation technique through time-delay overlapping method. After that, we call this modulation technique the time-shifting non-orthogonal modulation technology (TS-NMT). During the research of TS-NMT, we gradually realized that the wave forming method based on time-delay overlapping in TS-NMT is a practical method to break through the Shannon Limit.

### 4.1 The waveform structure of TS-NMT

Firstly, the waveform of the TS-NMT symbol constructed by the time-delay overlapping modulation technique is shown in Fig. 4. A single sinusoidal or cosine wave is called the base sub-carrier, and multiple subcarriers with identical frequency form a frequency sub-channel. It is similar with OFDM and they both belong to the multi-carrier category. There are many forms of TS-NMT and they can be uniformly expressed as follows,

$$\left. \begin{array}{l} g_t(t) = \sum_{h=1}^{H} a_h \bar{g}_t \omega_h (t - \tau_h) \\ g_r(t) = \sum_{h=1}^{H} a_h \bar{g}_r \omega_h (t - \tau_h) \end{array} \right\} t \in T, \tau_h \in T_{dh} \qquad (19)$$

where subscripts t and r represent the transmitter and the receiver, respectively. $T$ is the period of the symbol of the synthetic wave, and $T_{dh}=T_{d(h+1)}=T_d$ represents the duration of each subcarrier and is called the lifetime. $g_t(t)$ and $g_r(t)$ are the waveforms of the transmitted and received symbol, respectively. $a_h \bar{g}_t \omega_h (t - \tau_h)$ and $a_h \bar{g}_r \omega_h (t - \tau_h)$ are subcarriers, in which $a_h$ is the modulation amplitude and $\bar{g}_t \omega_h (t - \tau_h)$ and $\bar{g}_r \omega_h (t - \tau_h)$ are base-subcarriers. Base-subcarriers can be sinusoidal, cosine or other waveforms with no DC component, and $\omega_h$ is the angle frequency. It is worth noting that, $\omega_h = \gamma \omega_{h+1}$ and $\gamma$ can be any rational number, which means that orthogonality of frequency sub-channels is not needed. $\tau_h$ is the time delay, and the starting points of all the subcarriers are in the lifetime of the first subcarrier. Obviously, $\tau_0 = 0$. We can see that there are three differences between TS-NMT and OFDM:

**(1) There are time delays among each subcarrier;**

**(2) Orthogonality in the frequency domain is not required;**

**(3) The number of the sub-carriers in a single frequency sub-channel is not limited to 2. Specifically, there is no upper limit of the number of the sub-carriers.**

In addition, due to the interference of the channel fading and the noise, $\bar{g}_r \omega_h (t - \tau_h) \neq \bar{g}_t \omega_h (t - \tau_h)$ under normal circumstances. However, $\bar{g}_r \omega_h (t - \tau_h) \approx \bar{g}_t \omega_h (t - \tau_h)$ can be realized through channel estimation and equalization.

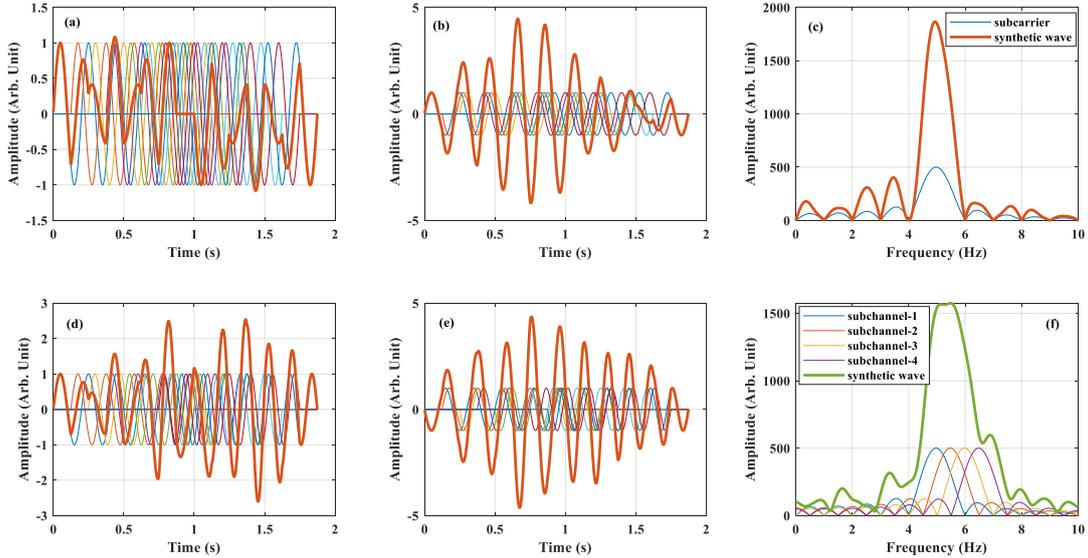

**Fig. 4.** Waveform structure of TS-NMT. (a) TS-NMT/S waveform structure in which all modulation amplitudes of base subcarriers are +1. (b) TS-NMT/S waveform structure in which the modulation amplitude of each base subcarrier is ±1 alternately. (c) Spectrum of a TS-NMT synthetic wave. (d) TS-NMT/F waveform structure. There are 4 frequency sub-channels, and each frequency sub-channel contains 2 base-subcarriers. All the modulation amplitudes are +1. (e) TS-NMT/F symbol. All the parameters are same with that in (d) except for the modulation amplitude of each base subcarrier is ±1 alternately. (f) Spectrum of each frequency sub-channel and the synthetic wave.

### 4.2 Demodulation methods of TS-NMT

The demodulation of TS-NMT can be realized through a specific method of solving the linear equation set. The diagram of an anatomic TS-NMT symbol with 4 subcarriers and the simple demonstration of the demodulation process are shown in Fig. 5. The goal of demodulation is to solve the modulation amplitude of each base-subcarrier from the periodic symbol waveform in turn. The modulation amplitude of a base-subcarrier can be obtained through the integral operation expressed as $\frac{1}{2}a_h = \int_{T_{dh}} a_h \sin\omega_h(t-\tau_h)\sin\omega_h(t-\tau_h)dt$, which is defined as the coherent operation. Due to the overlapping of all the subcarriers in $T_{d1}$, although it is impossible to distinguish the amplitudes of all the subcarriers through only one coherent operation, we can construct and solve an equation set to get all the modulation amplitudes.

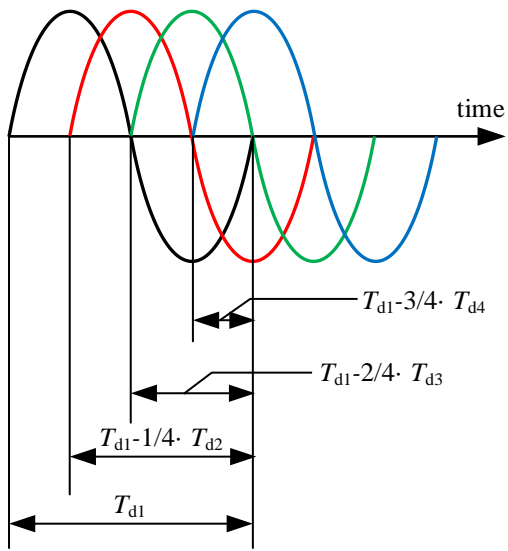

Interpretation of left picture:
During the demodulation, we use the first base-subcarrier to do the coherence operation with the received signal $g_r(t)$. This process can be divided into 4 parts, i.e. $T_{d1}$, $T_{d2}$, $T_{d3}$ and $T_{d4}$. Except for the first subcarrier, other subcarriers are all partial waveforms. By adding those coherent operation we can get an equation $r_{11}a_1+r_{12}a_2+r_{13}a_3+r_{14}a_4=B_1$. Using all base subcarriers to do coherent operations with the received signal in turn, We can get a system of linear equations $\boldsymbol{RA}=\boldsymbol{B}$.

**Fig. 5.** Diagram of the demodulation of TS-NMT.

In the linear equation set, $\boldsymbol{R}$ is the demodulation matrix, in which the elements are obtained through coherent operations between each base-subcarrier, i.e.

$R = [r_{hk}, h, k = 1, ..., H], r_{hk} = \int_{T_{dhk}} \bar{g}_r \omega_h (t - \tau_h) \bar{g}_r \omega_k (t - \tau_k) dt$. $A$ is the amplitude vector, and it is consisted of the quantified amplitude as $A = [a_h, h = 1, ..., H]$, which is the result of the linear equation set. It can be expressed as

$$A = R^{-1}B \tag{20}$$

where $B$ is the coherent vector and consists of the results of coherent operations, which is defined as $B_h = \int_{T_{dh}} g_r(t) \bar{g}_r \omega_h (t - \tau_h) dt, h = 1, ..., H$.

### 4.3 Transmission rate and performance analysis of TS-NMT

In [33], the theoretical analysis and numerical simulation results of the comparison between TS-NMT and OFDM is detailed, and the advantages of TS-NMT over OFDM is demonstrated, such as high spectral-efficiency, low peak-to-average power ratio (PAPR), strong anti-frequency shifting and low power consuming, etc.

Next, we will analyze the feasibility of TS-NMT as a practical method to break through the Shannon Limit. Four aspects are considered: power, transmission rate and efficiency, ill-condition of the equation set and the sampling frequency of the system.

Firstly, it can be seen from Fig. 4 intuitively that the synthetic wave of TS-NMT contains superposition of positive and negative amplitudes during some time sections, which makes the composite amplitude very small or even approximate zero. The numerical simulation results also indicate that the power of the whole synthetic wave doesn't increase proportionally with the number of subcarriers, which means that higher power efficiency can be achieved by TS-NMT. From Fig. 6(a), comparing with OFDM, the increment of the TS-NMT symbol power along with the number of subcarriers is much lower.

In the aspect of data transmission rate, the comparison of the spectral-efficiency between TS-NMT and the Shannon Limit is given in Fig. 6(b). We can see that the spectral-efficiency of TS-NMT exceeds the Shannon Limit, especially in the higher *SNR* domain. It is worth noting that, for TS-NMT/S, the frequency of all the subcarriers are the same, which means that the TS-NMT/S symbol occupies only one frequency sub-channel. Meanwhile, for TS-NMT/F, although multiple frequency sub-channels are occupied, due to the non-orthogonal of each frequency sub-channel, the spectral-efficiency is much higher than any existed multi-carrier modulation technologies, such as OFDM.

Thirdly, as we all know that, there is a problem of ill-condition in the numerical solution of the equation set, i.e. the perturbation of the equation set will bring large deviation between the accurate solution and the numerical solution. For TS-NMT, we proposed a unique method through adding additional wave to solve the ill-condition problem. From Fig. 6(c) and (d), we can see that the channel capacity can be further increased when there exists additional waves and the frequency of error is further decreased.

Last but not the least, a higher sampling frequency exceeding that regulated by the sampling theorem is required in TS-NMT, and the transmission rate is proportional to the sampling frequency. For this reason, some researchers consider that the sampling theorem employed by Shannon is not perfect. This paper did not draw that conclusion. In fact, according to Eq. (8) and the signal uncertainty principle, the continuous of the signal and $T \rightarrow \infty$ ensure the completeness of the sampling theorem under the frame of the signal uncertainty principle. We treat the enhancement of the sampling frequency only as a cost of the realization of TS-NMT.

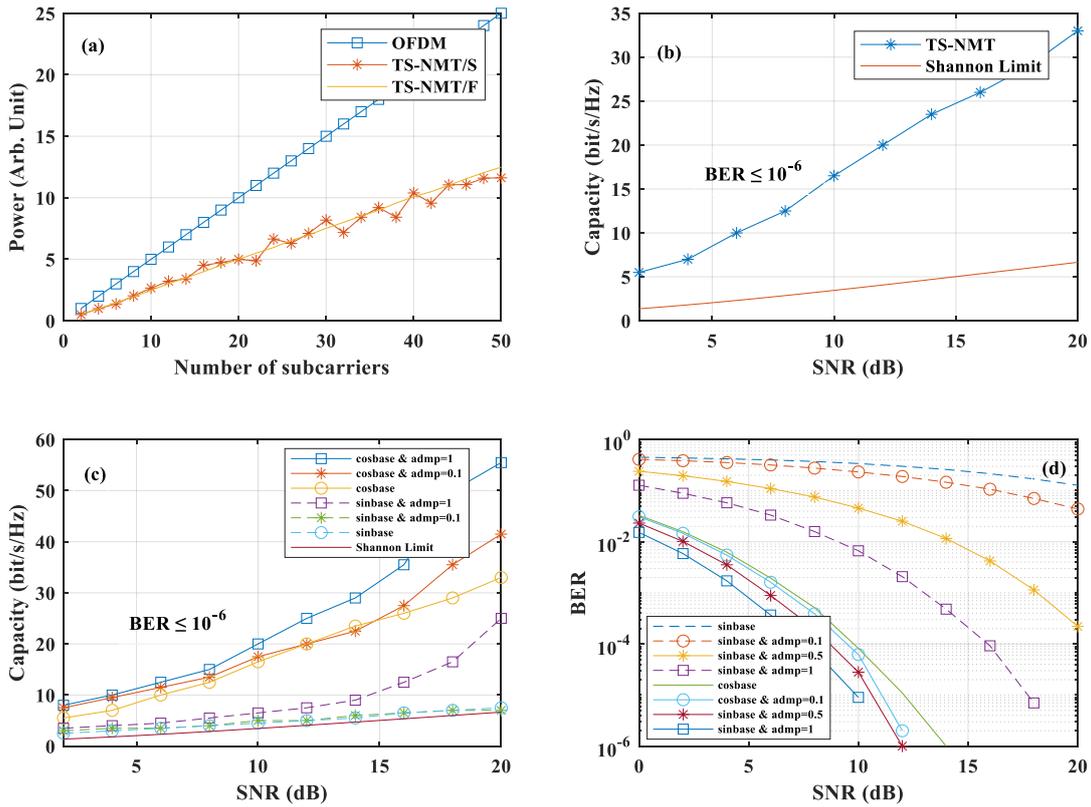

**Fig. 6.** Simulation results of the performance of TS-NMT. (a) Comparison of the symbol power of OFDM, TS-NMT/S and TS-NMT/F. (b) Channel capacity of TS-NMT. (c) Channel capacity of TS-NMT with additional waves. (d) BER performance of TS-NMT.

## Discussion

In the research of TS-NMT for more than 20 years, we gradually realize that the delay overlap method is an effective technical means to improve the transmission efficiency. For a long time, we have thought that this technology may break the limit of Shannon. The curve in Figure 7 shows that the slope of TS-NMT transmission efficiency curve is greater than that of Shannon efficiency curve, and the transmission rate will exceed Shannon limit when it is greater than a certain signal-to-noise ratio. This is a resoulute given by Dr. Cao Qisheng, a student of Liang Dequn, in his doctoral thesis [34], which has shown that the method contained in TS-NMT can break the Shannon limit. However, for the sake of prudence, we did not claim that breaking through Shannon limit can be broken.

In fact, we can interpret the efficiency of TS-NMT more intuitively. Through time-delay overlapping method, a single physical channel is divided into multiple logical channels, and the efficiency of the single physical channel is increased through addition of all the logical channels.

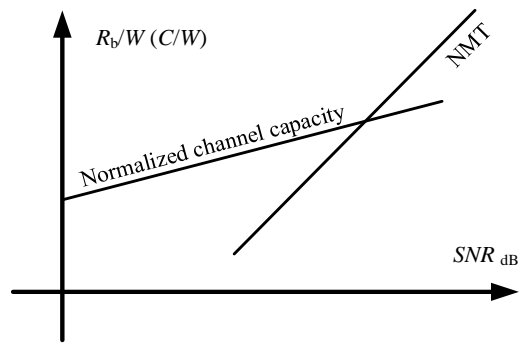

**Fig. 7.** Theoretical capacity of TS-NMT in [34]. The curve of normalized channel capacity is a simplifying version of the Shannon limit curve. The curve of TS-NMT is also a simplifying version of TS-NMT.

Through the study of breaking the Shannon limit, especially connecting the uncertainty principle in quantum mechanics with our research, we think that the research on information in the scientific community is not deep enough. Shannon only gives the measure of information, but does not answer what the physical essence of information. **We have following assumptions.** Up to now, people think that the world is made up of matter, which can be infinitely divided into smaller elementary particles. According to quantum mechanics, the states of these basic particles cannot be accurately measured. From a philosophical point of view, the author believes that the world is always uncertain, but human beings will constantly measure it. The author also believes that the world should be viewed in this way. She is composed of **hierarchical information clusters.** The unknown information cluster is called the original information cluster, which transmits information through the more dimensions the random codes. The number of dimensions will increase geometrically with the number of layers. Information clusters can be gathered or separated from each other. Information clusters with tight gathered and sufficient capacity are called matter. Capacity is measured by the complexity of the coding method. If we can find the way or code of information exchange between information clusters, we may have a deeper understanding of the world!

This paper discusses the Shannon channel capacity formula from the perspective of signal uncertainty principle to prove its completeness. In this process, it is found that Shannon's knowledge of information is only to quantify information, but still does not reveal the physical nature of information. The concept of hierarchical information group proposed by us may reveal the physical nature of information to a certain extent. Let's test the rationality of Darwinian evolution by discussing the controversial issues.

In Darwin's *Origin of Species*, he put forward a theory that organisms and organs are constantly developing. It also provides a test to test whether the theory is a pseudo Theory: "if we can prove the existence of any complex organ, they may not have been formed through numerous successive minor modifications, and my theory will definitely collapse." The main point of the argument is precisely from this. This can be translated into the question of "looking at the outside world from a discrete or continuous point of view". This is actually a manifestation of wave particle duality in quantum mechanics. Wave particle duality is also the ultimate feature of human cognitive ability.

In our previous idea of hierarchical information group, the distance between levels can be changed according to human cognitive ability and research needs. When it tends to zero, it is continuous, otherwise it is discrete. The evolution can be described as follows: the aggregation process of information clusters at any level is a random process with certain distribution. Among the new information clusters generated, some are relatively stable and some are unstable. Those stable information clusters are easy to be inherited to the higher level and combined with other information clusters. According to this point of view, we can explain the formation of complex organs in animals, which is an important example of Darwin's evolution being questioned, such as how the eyes are generated? We will only use the example of how the crystal composed of protein forms. It can be imagined that a large number of protein films with different thickness and density are formed in a certain layer, some of which have lens characteristics and are relatively stable information groups. When they are just combined with the information groups of some animals at a higher level, an animal with eyes is generated (note, for the sake of simplification, many details of eye structure are omitted here)!

Modern biology has used the concept of coding and corresponding technology in the research of genetic genes. And the coding we are talking about here is a lower level of information coding. The author believes that it is possible to prove the validity of this hypothesis if we look for the evidence of information coding in the chemical reactions from inorganic to organic. If this hypothesis is proved, it will open up a new way for us to understand the world.

## Acknowledgement

First of all, we would like to express our deep gratitude to Mr. Sun Changnian. Since 1997, he has provided professor Liang with the opportunity to explore efficient digital modulation methods. He was also the first one to support the initial idea of TS-NMT. When the rationality of TS-NMT was proved by simulation, we have been cooperating for a long time since 2005 with the goal of using TS-NMT for xDSL. Mr. Sun Changnian was responsible for the development of hardware prototype and commercial financing. Unfortunately, due to the lack of funds and human resources, we have not been able to enter the business road! However, a prototype for experiment has been made, which helps us to complete the verification of short-distance wired and wireless physical channels. We also thank three doctoral students and 33 master's students from Xi'an Jiaotong University and Dalian Maritime University who have done a lot of computer simulation and physical channel experiments from 1997 to 2010. Funding: This research was supported via the Natural Science Foundation of China (NSFC) (60272017, 60772160). Author contributions: Dequn Liang wrote the first draft and reviewed the final draft. Xinyu Dou completed the computer simulation and editing of the full text mentioned in the article. Competing interests: Dequn Liang is the author of two international patents related to this work (application. no. 00123342.4 and 200810119412.1). Xinyu Dou declares no competing interests. Data and materials availability: All data needed to evaluate the conclusions in the paper are present in the paper. Additional data related to this paper may be requested from the authors.